# Direct observation of the Yb(4f$^{13}$6s$^2$)F states and accurate determination of the YbF ionization energy

Thomas D. Persinger[a], Jiande Han[a], Anh T. Le[b], Timothy C. Steimle[c] and Michael C. Heaven[a†]

a.  Department of Chemistry, Emory University, Atlanta, GA 30322

b.  School of Chemistry and Biochemistry, Georgia Institute of Technology, Atlanta, GA, 30318

c.  School of Molecular Sciences, Arizona State University, Tempe, AZ 85287

**Abstract**

YbF has been identified as a molecule that can be used to investigate charge-parity symmetry violations that are beyond the Standard Model of particle physics. Cooling to sub-milli-Kelvin is advantageous for experiments that probe manifestations of these symmetry violations. One approach involves laser cooling of YbF via the A$^2\Pi_{1/2}$-X$^2\Sigma^+$ transition. However, it appears that cooling by means of this transition may be limited by the radiative loss of population from the cooling cycle. YbF has low-energy states that arise from the Yb$^+$(4f$^{13}$6s$^2$)F$^-$ configuration. Recent theoretical calculations predict (Zhang et al J. Mol. Spectrsc. 386 11625 (2022)) that radiative decay from A$^2\Pi_{1/2}$ to the 4f$^{13}$6s$^2$ states occurs with a branching fraction of approximately 10$^{-3}$. In the present study we have used dispersed laser induced fluorescence spectroscopy to the observe the lowest energy 4f$^{13}$6s$^2$ states. These measurements were carried out using excitation of previously unobserved YbF transitions in the near UV spectral range. An accurate ionization energy (IE) for YbF is also reported. A two-color photoionization technique was used to determine the IE and observe the v$^+$=0-3 vibrational levels of YbF$^+$ X$^1\Sigma^+$.

†Corresponding author. E-mail, mheaven@emory.edu. Phone, (404) 727 6617

**Introduction**

Here we present experimental identification of low-lying electronic state of YbF using dispersed laser induced fluorescence (DLIF), which has implications for photon cycling and laser cooling experiments. We also accurately determine the ionization energy (IE) and identify the observe the $v^+$=0-3 vibrational levels of YbF$^+$ X$^1\Sigma^+$ which will assist in new detection techniques. The heavy polar diatomic molecule YbF has received considerable attention over the past 25 years since it was first proposed as a venue for electric dipole moment (EDM) measurements [1-3]. It was noted early on [1] that the $^{174}$YbF isotopologue is well suited as a venue for determination of the leptonic sector electron EDM (eEDM, $d_e$), and that the $^{173}$YbF isotopologue can be used to investigate both $d_e$ and the nuclear magnetic quadrupole moment (MQM) of the $^{173}$Yb(I=5/2) nucleus (hadronic sector). The MQM of the $^{173}$Yb nucleus arises from the composite effect of the EDM of the constituent nucleons, $d_N$, and the intranuclear $T$, $P$-odd forces (where $T$ and $P$ indicate time-reversal and discrete parity symmetries). The $T$, $P$-odd MQM interacts with the gradient of the magnetic field produced of the unpaired electron to produce a molecular EDM [4-6]. An EDM of an elementary particle (e.g. $d_e$ and $d_N$) violates both $T$, and $P$ symmetries. The Standard Model (SM) predicts that EDMs are exceptionally small (e.g. $d_e$ (SM) less than $10^{-38}$ $e$ cm) whereas theories beyond the Standard Model (BSM) frequently predict much larger values which in turn, due to the enhancement effect of the polar molecule, may produce a molecular EDM large enough to be measured by precision spectroscopy.

The two reported YbF-based attempted EDM measurements [7-9], both performed at Imperial College-London, have focused on the determination of $d_e$. The goal is to observe the $d_e$-induced shift of the spacing of the $\left| N=0, F=1; M_F=+1 \right\rangle$ and $\left| N=0, F=1; M_F=-1 \right\rangle$ levels of the $X^2\Sigma^+(v=0)$ vibronic state. The small $T$, $P$-odd energy shifts in YbF are determined by simultaneously subjecting the molecules to parallel electric and magnetic fields and using the measured spin-precession angular frequency [10,11] to determine the energy difference of states of opposite angular momentum projection. The $A^2\Pi_{1/2} - X^2\Sigma^+(0,0)$ band is used for state preparation and detection. The uncertainty of the extracted estimate for $d_e$ is inversely proportional to the coherence time, $\tau$, and, in the shot-noise limit, the square root of the photon counts of the $A^2\Pi_{1/2} - X^2\Sigma^+(0,0)$ LIF signal used for detection[11]: $\delta(d_e) \propto \dfrac{1}{\tau\sqrt{counts}}$. The coherence time is dependent upon the flight time and divergence of a molecular beam of YbF as it travels through the parallel fields. The first YbF-based eEDM measurement [8] used a beam produced by skimming the output of a high temperature effusive oven source and established an upper limit for $d_e$ of $(-0.2 \pm 3.2) \times 10^{-26}$ $e$ cm. The second reported YbF-based eEDM measurement attempt used a laser ablation supersonic expansion beam source and established an upper limit for $d_e$ of $(-2.4 \pm 5.7_{stat} \pm 1.5_{syst}) \times 10^{-28}$ $e$ cm.

A laser cooled sample could drastically improve the coherence time. One-dimensional [12] and two-dimensional [13] translational cooling of YbF has been recently demonstrated using a cryogenic buffer gas source and photon cycling of the $P_1(1)$ and $Q_{12}(1)$ lines of the $A^2\Pi_{1/2} - X^2\Sigma^+(0,0)$ band for the main laser cooling transitions. Modelling the 2D cooling [13] suggested that there was a leak out of the cooling cycle to unobserved low-lying states having a dominant $Yb^+(4f^{13}(^2F_{7/2})6s^2)F^-$ configuration. These unobserved low-lying excited states had been used of explain the observed anomalous sign [2] (relative to the alkaline earth monohalides) of the spin-rotation parameter, $\gamma$, for the $X^2\Sigma^+(v=0)$ and the observed strong vibrational dependence of this parameter [14].

A nearly complete review of the previous spectroscopic investigations of YbF can be found in recent a de-perturbation analysis [15]. That study addressed the nature of the well characterized $A^2\Pi_{1/2}$ ($v$=0), [557], [561] states, and five other excited electronic states that have been observed at low spectral resolution. The results of relativistic electronic structure calculations for the 4f-hole states and the strengths of transitions between these states and the $A^2\Pi_{1/2}$ ($Yb^+(4f^{14}6p^1)F^-$) and $X^2\Sigma^+$ ($Yb^+(4f^{14}6s^1)F^-$)) states were also presented. The lowest energy 4f-hole excited states are derived from the $Yb^+(4f^{13}(^2F_{7/2})6s^2)F^-$ configuration where, under the influence of the ligand field, the atomic ion $J_a$=7/2 state splits into molecular states with $\Omega$=1/2, 3/2, 5/2 and 7/2, where $\Omega$ is the unsigned projection of the electronic angular momentum along the interatomic axis. The higher energy excited states are derived from the $Yb^+(4f^{13}(^2F_{5/2})6s^2)F^-$ configuration and the atomic ion $J_a$=5/2 state splits into molecular states with $\Omega$=1/2, 3/2, and 5/2. The energy interval between the $J_a$=7/2 and 5/2 manifolds is expected to be approximately 10,000 cm$^{-1}$ based upon the energies of the $^2F_{7/2}$(E=21,418 cm$^{-1}$) and $^2F_{5/2}$ (E=31,568 cm$^{-1}$) states arising from the [Xe]$4f^{13}6s^2$ configuration of Yb$^+$. The electronic structure calculation, combined with the experimentally determined molecular constants and de-perturbation analysis [15] , puts the three $J_a$ =5/2 states in the range 18,000 cm$^{-1}$ -21,000 cm$^{-1}$ and the five $J_a$ =5/2 states in the range 8,000 cm$^{-1}$ -11,000 cm$^{-1}$. The nearly degenerate [561] (E~561 THz =18713 cm$^{-1}$) and [557] (E~557 THz =18580 cm$^{-1}$) states were assigned [15] as admixtures of $A^2\Pi_{1/2}(v=1)$ and the $\Omega$=1/2 state of the Yb$^+$(4f$^{13}$($^2$F$_{5/2}$)6s$^2$)F$^-$ manifold.

Given this background, it is of interest to carry out experimental tests of the theoretical predictions. In the present study we have focused on determination of the energies of the 4f$^{13}$($^2$F$_{7/2}$)6s$^2$ states. The strategy for these measurements was to observe transitions down to these states following laser excitation of suitable electronically excited states. The previously characterized A$^2\Pi$ and B$^2\Sigma^+$ states [14,16-18] were not considered viable upper state candidates as the transitions from these states to the 4f$^{13}$6s$^2$ states were predicted [15] to be weak, and the long wavelengths of the emissions would not be compatible with the detectors available for this project. To circumvent these problems, we considered it likely that there would be higher energy states that would have mixed configurational parentage. This would permit both excitation from the ground state and emission back to the 4f$^{13}$6s$^2$ states. A search for such states was made in the 31000 – 33500 cm$^{-1}$ range, as this would result in emissions down to the 4f$^{13}$($^2$F$_{7/2}$)6s$^2$ states at wavelengths that could be detected using standard, high-gain photomultiplier tubes. As described in the following, this proved to be a successful approach.

The apparatus used for these measurements was also configured for accurate determinations of ionization energies (IE's), and the IE of YbF is a property of both thermodynamic significance and theoretical interest [19,20]. Uncertainty in the IE propagates through schemes where measurements of the bond dissociation energy for YbF$^+$ are used to assess the bond dissociation energy for the neutral molecule[19-24]. The previous best estimate [19,24] for the IE of YbF was 47700±400 cm$^{-1}$. In the present work we have used resonant two-photon ionization techniques to obtain an accurate IE for YbF, and to determine the vibrational constants of YbF$^+$, X$^1\Sigma^+$.

**Experimental**

The instrumentation used to record gas-phase spectra for YbF has been described previously [25]. YbF was obtained by pulsed 1064 nm Nd/YAG laser ablation of the surface of a Yb rod. The rod was mounted in a Smalley-type jet expansion source [26] where it was rotated and translated to avoid pitting. The carrier gas was a mixture of 0.5% SF$_6$ in He, supplied by a pulsed valve (Parker-Hannifin series 9) at a source pressure of 5 atm. Laser induced fluorescence (LIF) and dispersed laser induced fluorescence (DLIF) spectra were recorded with the excitation laser beam set to cross the jet expansion approximately 7.5 cm downstream from the nozzle orifice. LIF was collected along an axis that was perpendicular to both the laser beam axis and the jet expansion axis. For the recording of LIF data, a long-pass filter was used to reduce the scattered laser light, and the filtered fluorescence was detected by a photomultiplier tube (Photonis XP2020). DLIF spectra were obtained when the long-pass filter was replaced by a 0.25 m monochromator (Jarrell-Ash 82-410).

Two tunable dye laser systems were used in these experiments. These were Nd/YAG pumped systems, consisting of a Lambda Physik Scan Mate Pro driven by a Quantel Q-smart 850 Nd/YAG, and a Continuum ND6000 dye laser driven by a Powerlite 8000 Nd/YAG laser. The dye lasers operated with linewidths (FWHM) of approximately $0.3$ cm$^{-1}$. Frequency-doubling of the output from the ND6000 dye laser was used to generate light in the 31000-33500 cm$^{-1}$ range. Wavelength calibration of the dye lasers was established using a Bristol Instruments model 821 wavemeter. The wavelength of the laser fundamental was measured to calibrate the frequency doubled light. Fluorescence decay curves were acquired using a digital oscilloscope (LeCroy WaveSurfer 24Xs) to signal average 256 laser pulses per trace. The monochromator was used to isolate the kinetics associated with specific, well-resolved emission lines.

Resonantly enhanced two-photon ionization (RE2PI), photo-ionization efficiency (PIE) and pulsed field ionization- zero kinetic energy electron (PFI-ZEKE) measurements were carried out in a differentially-pumped vacuum chamber that was equipped for time-of-flight mass spectrometry and electron detection [25]. RE2PI and PIE spectra were recorded with mass-resolved cation detection. RE2PI data for transitions occurring in the near-UV spectral range were obtained using one-color, two-photon ionization. This was possible because the excited state energies were more than half of the IE. RE2PI spectra for lower energy states were observed using two-color ionization. All two-color measurements were carried out using spatially overlapped, counter-propagating laser beams. Digital delay generators (Stanford Research Systems, DG 535) were used to synchronize the light pulses.

PIE curves were recorded with the first laser set to populate an excited state of YbF. The wavelength of the second laser was then swept to locate the onset of ionization. These measurements were conducted with the ionization zone located between the charged electrodes of the mass spectrometer. The local electric field of 364 V cm[-1] caused a depression of the IE by 115 cm[-1]. PFI-ZEKE spectra for YbF[+] were recorded with sequential laser excitation occurring under nominally field-free conditions. Sequential excitation to long-lived Rydberg states was followed by the application of a 1.4 V cm[-1] pulse to induce field ionization and acceleration of the electrons to a microchannel plate detector. The delay between the laser excitation and ionizing field pulses was 2 µs.

**Results**

The first step for this project was to locate electronically excited levels of YbF in the near-UV spectral range. The RE2PI technique was used for this search as the mass-resolved ion detection ensured that the carrier of the observed features would be YbF. Figures 1 and 2 show examples of intense transitions that were found in the target energy range. We did not attempt to rotationally resolve these features, but the band contour can be discerned in Fig.1. The relatively sharp Q-branch, with a hint of blue shading, indicates that the rotational constant had increased slightly on electronic excitation. This change also results in a P-branch band head that is approximately 16 cm[-1] below the Q-branch maximum. Due to spectral congestion, it is more difficult to interpret the band contours of Fig. 2. For the present purpose, we note that the Q-branch of Fig. 1 (at 31052 cm[-1]), and the strong features at 32795 cm[-1] and 33058 cm[-1] of Fig. 2 (as indicated by arrows) were used in the recording of DLIF spectra.

Figure 3 shows an expanded segment of the DLIF spectrum obtained by exciting the 31052 cm[-1] band. The only other feature in this spectrum was a weak emission back to the ground state v"=0 level (resonance emission). This had an intensity that was approximately 1/10 of the strongest feature (without correction for the wavelength dependent sensitivity of the detection system). The emission bands of Fig. 3 were easily assigned using the $4f^{13}6s^2$ energy levels predicted by electronic structure calculations [27-29]. Transitions to the $\Omega$"=1/2 and 3/2 states were present, but emission down to $\Omega$"=5/2 levels were not observed. Based on the $\Delta\Omega$=0, ±1 selection rule, we conclude that the upper state is $\Omega$'=1/2. In the following, we label the UV excited states using the notation $[T_0/1000]\Omega$, where $T_0$ is the term energy in units of cm[-1]. Hence, the upper state label for the band shown in Fig. 1 is [31.05]1/2.

Figure 4 shows the DLIF spectrum obtained by exciting the UV band at 32795 cm[-1]. The energy scale for this trace is the excitation energy minus the emission energy, giving the energy of the lower state relative to the ground state zero-point level. There are three interesting details of this spectrum. First, as was the case for Fig. 3, the resonant emission feature was weak (not shown). The second point of interest is that the most intense feature is the emission down to the $\Omega''$=5/2, v''=0, indicative of $\Omega'$=3/2. Lastly, the lines at 11721 and 14689 cm$^{-1}$ do not originate from the initially excited level. They are the B$^2\Sigma^+$-X$^2\Sigma^+$ and A$^2\Pi_{1/2}$- X$^2\Sigma^+$ 0-0 transitions, where the upper states have been populated by radiative relaxation from [32.80]3/2.

The DLIF spectrum obtained by exciting at 33058 cm$^{-1}$ was dominated by transitions down to the $\Omega''$=1/2 and 3/2 states, indicating $\Omega'$=1/2. The B$^2\Sigma^+$-X$^2\Sigma^+$ and A$^2\Pi$- X$^2\Sigma^+$ 0-0 emission bands were also present, but for the [33.06]1/2 upper state, the A$^2\Pi_{3/2}$ state was preferentially populated over A$^2\Pi_{1/2}$.

Energies for the 4f$^{13}$($^2$F$_{7/2}$) 6s$^2$ $\Omega''$=1/2, 3/2 and 5/2 states, derived from the DLIF spectra, are listed in Table 1. Short vibrational progressions were observed, but the resolution was not sufficient for determination of the vibrational anharmonicity constants. The average vibrational intervals were 610, 607, and 610 cm$^{-1}$ for $\Omega''$=1/2, 3/2 and 5/2, respectively.

To examine the radiative decay kinetics of the UV excited states, we recorded fluorescence decay curves obtained by isolating emission features of interest. Fig. 5 shows data obtained by monitoring the emissions from the [33.06]1/2, B$^2\Sigma^+$, v'=0 and A$^2\Pi_{3/2}$, v'=0 levels. Fitting a single exponential to the decay segment of the [33.06]1/2 state emission yielded a lifetime of 32(5) ns. The temporal traces for the B$^2\Sigma^+$ and A$^2\Pi_{3/2}$ state emissions show maxima that occur later than the maximum of the [33.06]1/2 emission. The shift was consistent with population of the A$^2\Pi_{3/2}$ and B$^2\Sigma^+$ states by radiative decay (the collision frequency in the downstream region of the jet-expansion was far too low to contribute via non-radiative population transfer). Fitting a radiative cascade model to the temporal data for the B$^2\Sigma^+$ and A$^2\Pi_{3/2}$ states defined fluorescence decay lifetimes of 50(5) and 35(5) ns, respectively. A lifetime for A$^2\Pi_{1/2}$ v'=0 of 28(2) ns was reported previously by Zhuang et al.[29]

Sequential two-photon ionization of YbF was examined using the A$^2\Pi_{1/2}$ – X$^2\Sigma^+$ 0-0 band at 18106.3 cm$^{-1}$ for the first excitation step. A preliminary value for the ionization threshold was then obtained by scanning the wavelength of the second laser over the range where the sum of the photon energies would cross the threshold for ionization (PIE scan). This process was monitored via the mass-resolved detection of $^{174}$YbF$^+$ ions. After correcting for the depression of the observed threshold due to the static electric field of the mass spectrometer, the threshold was found at 48,700(20) cm$^{-1}$.

The IE was refined by switching to the PFI-ZEKE technique, where ionization occurs under nominally field-free conditions. Fig. 6 shows the PFI-ZEKE spectrum for ionization to the YbF$^+$ X$^1\Sigma^+$, v$^+$=0 level. The structure of this band is consistent with ionization from the range of rotational levels populated by the initial excitation of the A$^2\Pi_{1/2}$ – X$^2\Sigma^+$ Q-branch maximum. The red-most edge of the band in Fig. 6 corresponds to ionization from the highest intermediate J-levels populated by the first laser pulse. As the energy of the second photon is increased, the ionization signal increases as lower J-levels are sequentially accessed. The cut-off on the blue side of the band occurs when the lowest energy J-level can be ionized. Hence, the blue-edge from Fig. 6 was used to define an IE of 48703(5) cm$^{-1}$. Scanning to higher energies revealed similar spectral features for the v$^+$=1, 2, and 3 levels. The energies of the YbF$^+$ states, relative to the YbF X$^2\Sigma^+$ v"=0 ground state, and YbF$^+$ X$^1\Sigma^+$ v$^+$=0, are listed in Table 2. Fitting to the Morse expression for vibrational energies yielded constants of $\omega_e^+$=604.9 and $\omega_e x_e^+$=2.7 cm$^{-1}$.

**Discussion**

Previous spectroscopic studies of YbF have examined transitions occurring at energies below 28000 cm$^{-1}$, so the near UV transitions used in this study are reported for the first time. The DLIF spectra show short vibrational progressions with intensity envelopes indicating that the upper state vibrational levels are v'=0, and that the molecular constants for the states are close to those of the low-energy 4f$^{13}$6s$^2$ states.

Pototschnig et al.[28] used relativistic electronic structure calculations to predict states of YbF with term energies up to 25023 cm$^{-1}$. Due to the large number of excited states, they were not able to assign the bands reported by Uttam and Joshi [30] for the 21700-2800 cm$^{-1}$ spectral range. Given the complexity at even higher energies, we can only provide a qualitative comment concerning the configurational parentage of the UV-excited states used in this work. The marked preference for emission back down to the 4f$^{13}$6s$^2$ states and the fact that initial electronic excitation had a small effect on the bonding suggests that the atomic ion parentage contains a substantial contribution from 4f$^{13}$5d6s (see Fig. 2 of ref. [28]).

A primary objective of the present study was to determine the energies of the low-energy 4f$^{13}$6s$^2$ states, and use this information to evaluate the quality of the predictions made in the work of Zhang et al.[15]  Table 3 compares the experimental results to the predictions of high-level relativistic electronic structure calculations.  Here it is apparent that the empirically adjusted calculations of Zhang et al.[15] yielded term energies that are close to the measured values (errors below 100 cm$^{-1}$).  Good agreement was also achieved for the harmonic vibrational constants.

The computational techniques used by Zhang et al.[15]  predicted the $4f^{14}6p$ and $4f^{13}6s^2$ states by means of single excitations from the ground state $4f^{14}6s$ configuration. The accuracy was not sufficient to capture the experimentally observed near degeneracy of the $A^2\Pi_{1/2}$ (v=1) state with the $4f^{13}(^2F_{5/2})6s^2$ $\Omega$=1/2 (v=0) state. The relative energies of the $4f^{13}(^2F_{5/2})6s^2$ and $4f^{13}(^2F_{7/2})6s^2$ manifold of states were fixed to the relativistic ab initio values. The absolute energies were empirically shifted to match the predicted energy (=18586 cm$^{-1}$) for the $4f^{13}(^2F_{5/2})6s^2$ $\Omega$=1/2 (v=0) state which was obtained from the depurturbation analysis of the experimentally observed energies of the eight vibronic states mentioned above. The agreement with the observed energy ordering and energy intervals for the $4f^{13}(^2F_{7/2})6s^2$ states strengthens the credibility of the transition moment calculations, which included values for the $A^2\Pi_{1/2}$-$X^2\Sigma^+$ vibronic bands that are critically important for the design of laser cooling schemes.  Table 3 also contains previous theoretical results obtained using high-level, relativistic methods.  Overall, the agreement between the calculations and the experimental results is respectable.

The previous best estimate [21,22,31] for the IE of YbF was 47700(400) cm$^{-1}$, based primarily on the study of Belyaev et al.[21]  The value measured by the PFI-ZEKE technique is higher by 1000 cm$^{-1}$.  The refined value for IE(YbF), combined with IE(Yb)=50443 cm$^{-1}$, shows that the bond dissociation energy of the ion is slightly greater than that of the neutral molecule (IE(Yb)-IE(YbF)=$D_0^+ - D_0$=1740 cm$^{-1}$).  Parker et al.[20] recently published a lower bound for the bond dissociation energy for YbF$^+$ of $D_0^+$>43885 cm$^{-1}$.  With the new value for IE(YbF) this translates to an upper bound for the dissociation energy for neutral YbF of $D_0$>42145 cm$^{-1}$.  This is larger than the estimate of 39000 cm$^{-1}$ by Barrow and Chojnicki [16], but the difference is not surprising given that their estimate was based on a long extrapolation of spectroscopic data.

Pototschnig et al.[28] predicted the IE for YbF as a side-issue in their theoretical study of the neutral molecule excited states.  Their method used the closed-shell species YbF$^+$ and YbF$^-$ to generate sets of starting orbitals.  Their best estimate for IE(YbF), obtained by extrapolation to the complete basis set limit, was 49901 cm$^{-1}$.  The small change in the dissociation energy that accompanied ionization ($D_0^+ - D_0$=1740 cm$^{-1}$) was consistent with the removal of the marginally anti-bonding 6s electron.

**Summary**

Dispersed laser induced fluorescence measurements have been used to examine the lowest energy states of YbF that correlate with the Yb$^+$(4f$^{13}$($^2$F$_{7/2}$)6s$^2$)F$^-$ electronic configuration. States with electronic angular momentum values of $\Omega$"=1/2, 3/2 and 5/2 were observed. The term energies for these states and their vibrational intervals were consistent with recent high-level computational predictions. Of particular interest, the experimental data support recent calculations that have been used to assess the potential for laser cooling of YbF via the A$^2\Pi_{1/2}$ – X$^2\Sigma^+$ transitions. These calculations indicate that re-pump lasers may be needed to compensate for radiative loss from the A$^2\Pi_{1/2}$ vibronic states to the 4f$^{13}$6s$^2$ states.

An accurate ionization energy for YbF was determined using the pulsed field ionization – zero kinetic energy photoelectron technique. The value obtained, 48703(5) cm$^{-1}$, was approximately 1000 cm$^{-1}$ higher than previous estimates. It is 1740 cm$^{-1}$ below the IE for Yb, which indicates that the bond dissociation energy of the ion exceeds that of the neutral by the same amount. A vibrational progression was observed for YbF$^+$ X$^1\Sigma^+$, providing the first experimental determination of the vibrational constants for the ion.

**Acknowledgment**

This work was supported by the National Science Foundation under grant CHE-1900555. We thank Prof. Lan Cheng (Johns Hopkins University) for, his helpful advice concerning the previous electronic structure calculations for YbF, and to Dr. Michael R. Tarbutt (Imperial College, London) for helpful discussion concerning the laser cooling of YbF.

Table 1.  Vibronic energies for the low-energy Yb$^+$(4f$^{13}$($^2$F$_{7/2}$)6s$^2$)F$^-$ states.

| v | $\Omega$=1/2 | $\Omega$=3/2 | $\Omega$=5/2 |
|---|---|---|---|
| 0 | 8470 | 9020 | 9810 |
| 1 | 9080 | 9630 | 10420 |
| 2 | 9690 | 10230 | |
| 3 | 10300 | 10840 | |

Energies in cm$^{-1}$ units. Errors are $\pm$10 cm$^{-1}$

Table 2.  Energies of the YbF$^+$ X$^1\Sigma^+$ vibrational states, relative to the zero-point levels of YbF X$^2\Sigma^+$ and YbF$^+$ X$^1\Sigma^+$

| v$^+$ | X$^1\Sigma^+$ (cm$^{-1}$) | T$_0$ (cm$^{-1}$) |
|---|---|---|
| 0 | 48703.0 | 0 |
| 1 | 49303.5 | 600.5 |
| 2 | 49897.0 | 1194.0 |
| 3 | 50489.8 | 1786.8 |

Table 3.  Comparison of observed and calculated energy levels for the low-energy Yb$^+$(4f$^{13}$($^2$F$_{7/2}$)6s$^2$)F$^-$ states.

| $\Omega$ | T$_0$(exp) | T$^a$ | T$^b$ | T$^c$ |
|---|---|---|---|---|
| 1/2 | 8470 | 8551 | 9627 | 6099 |
| 3/2 | 9020 | 9061 | 10180 | 6572 |
| 5/2 | 9810 | 9907 | 10967 | 7120 |

Energies in cm$^{-1}$ units.

*a*. Zhang et al. (ref. [15])  X2CAMF-EOM-CCSD calculations, empirically corrected.

*b*. Pototschnig et al. (ref. [28]) IHFS-CCSD, extrapolated to the complete basis set limit.

*c*. Dolg et al. (ref. [27]) Relativistic multi-reference configuration interaction calculations.

# Figure captions

1. One-color RE2PI spectrum of the YbF [31.05]1/2 – $X^2\Sigma^+$ transition. This trace was recorded using detection of the [174]YbF$^+$ isotopologue.

2. One-color RE2PI spectrum of YbF showing the excitation features used in recording DLIF spectra. This trace was recorded using detection of the [174]YbF$^+$ isotopologue.

3. DLIF spectrum obtained using excitation of the YbF [31.05]1/2 – $X^2\Sigma^+$ transition at 31052 cm$^{-1}$

4. DLIF spectrum obtained using excitation of the YbF [32.80]3/2 – $X^2\Sigma^+$ transition at 32795 cm$^{-1}$

5. Fluorescence decay curves recorded using excitation of the YbF[33.05]1/2- $X^2\Sigma^+$ transition at 33058 cm$^{-1}$

6. PFI-ZEKE spectrum of the YbF$^+$ $X^1\Sigma^+$ origin band. The first photon was set to excite the YbF $A^2\Pi_{1/2}$-$X^2\Sigma^+$ 0-0 band at 18106.3 cm$^{-1}$. The high-frequency oscillations of this trace were caused by shot-to-shot amplitude noise.

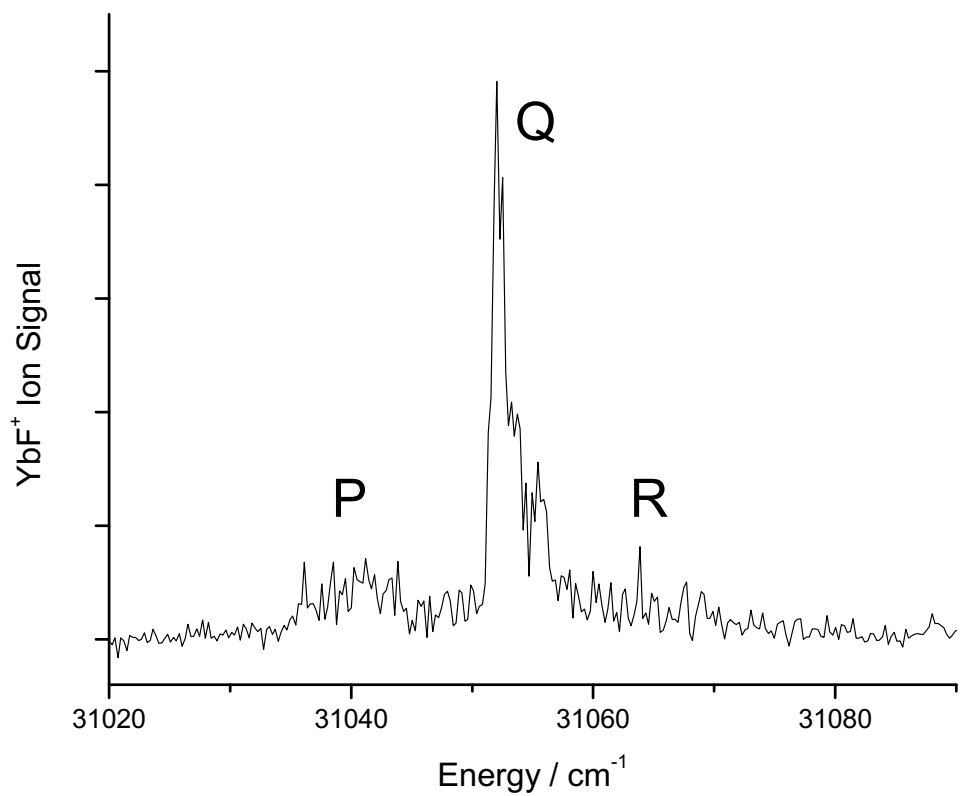

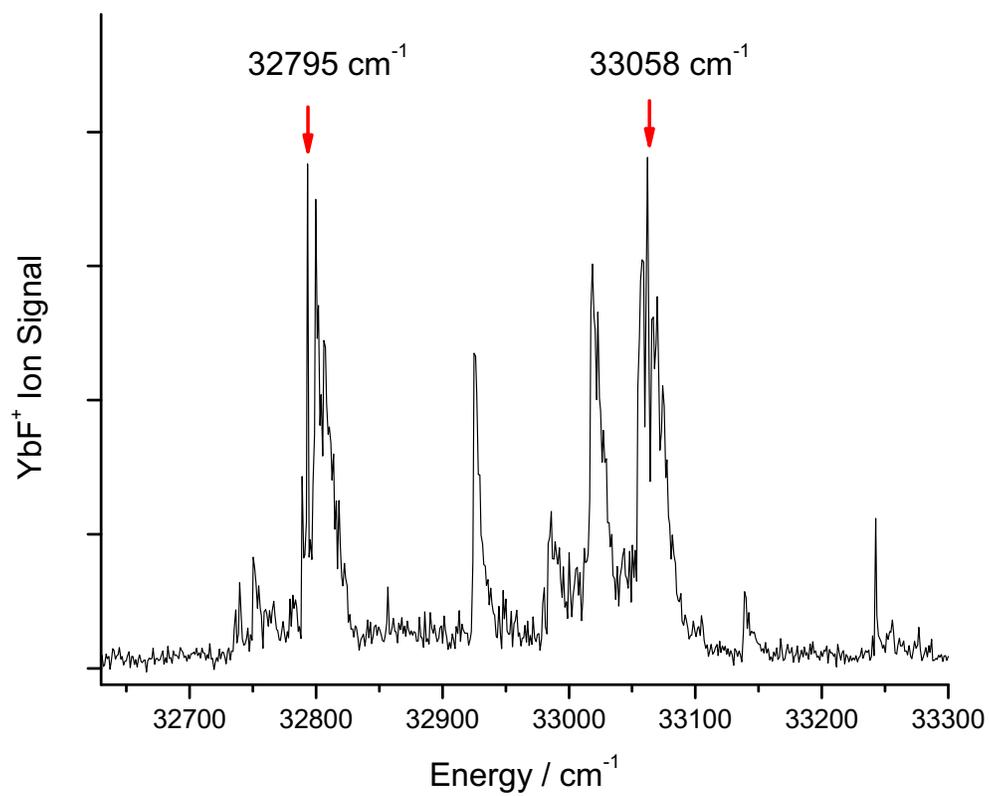

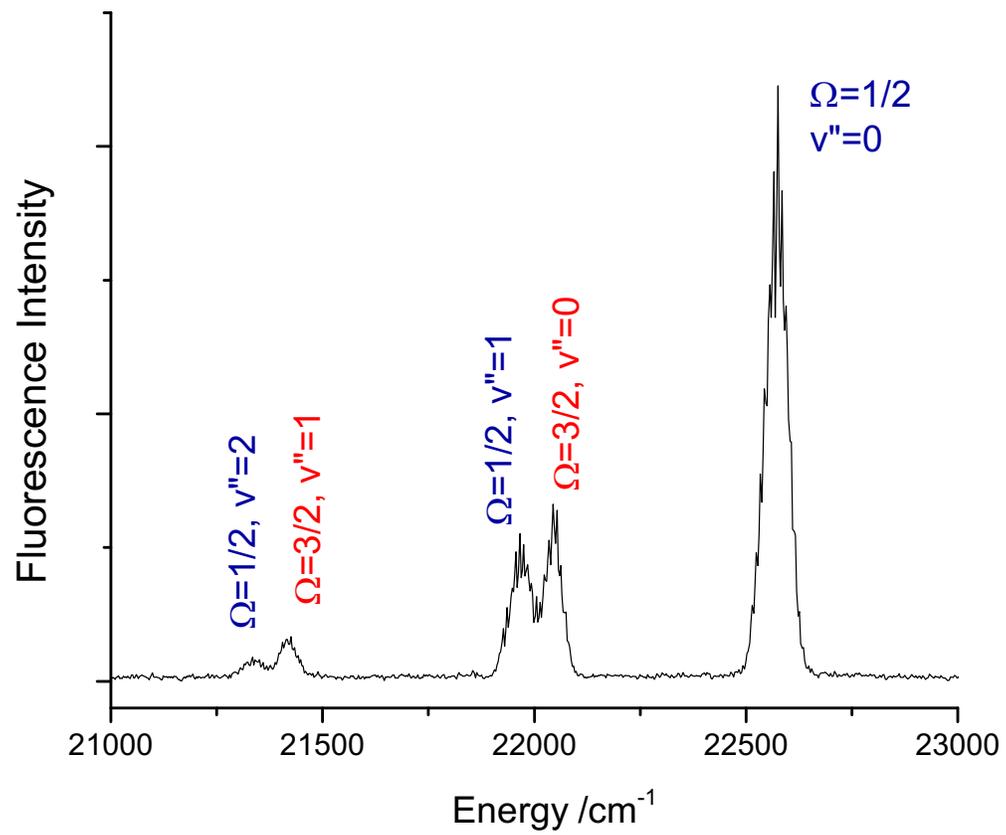

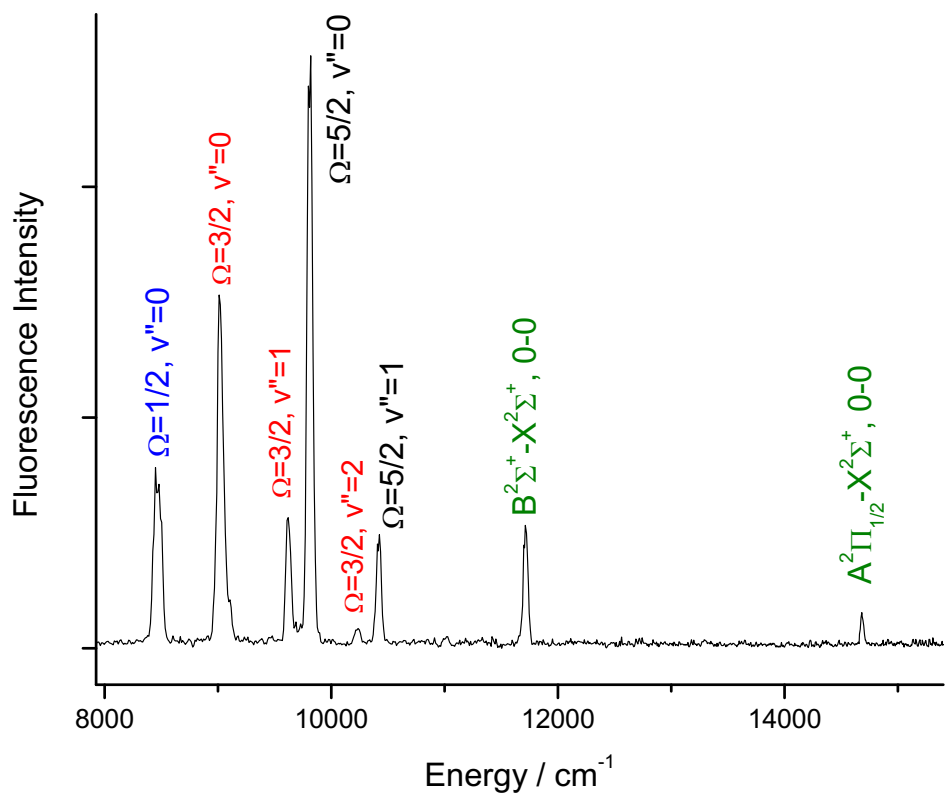

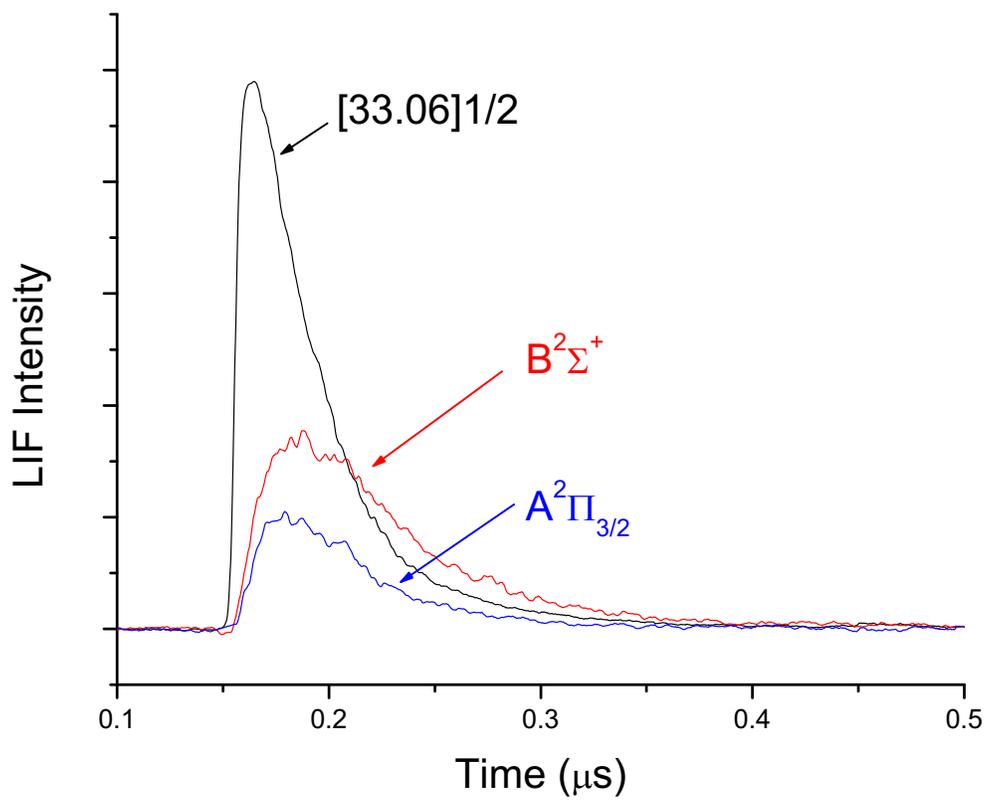

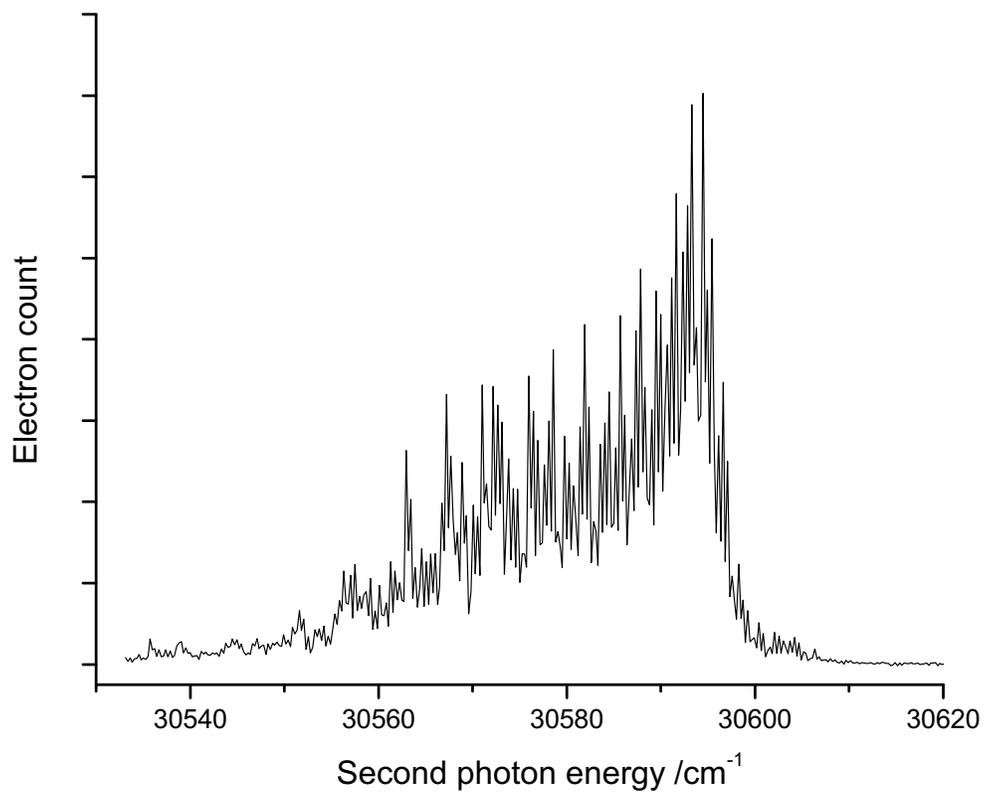